\documentclass[11pt]{article}\textwidth 6.5in\textheight 9in
\usepackage{amssymb}\usepackage[colorlinks]{hyperref}\usepackage{color}
	\usepackage{stmaryrd}\usepackage{mathrsfs}
	\usepackage{graphicx}
	\usepackage{amsmath}
\usepackage{caption}

\begin{document}
\setlength{\captionmargin}{27pt}
\newcommand\hreff[1]{\href {http://#1} {\small http://#1}}
\newcommand\trm[1]{{\bf\em #1}} \newcommand\emm[1]{{\ensuremath{#1}}}
\newcommand\prf{\paragraph{Proof.}}\newcommand\qed{\hfill\emm\blacksquare}

\newtheorem{thr}{Theorem} 
\newtheorem{lmm}{Lemma}
\newtheorem{cor}{Corollary}
\newtheorem{con}{Conjecture} 
\newtheorem{prp}{Proposition}

\newtheorem{blk}{Block}
\newtheorem{dff}{Definition}
\newtheorem{asm}{Assumption}
\newtheorem{rmk}{Remark}
\newtheorem{clm}{Claim}
\newtheorem{example}{Example}

\newcommand{\ab}{a\!b}
\newcommand{\yx}{y\!x}
\newcommand{\yux}{y\!\underline{x}}

\newcommand\floor[1]{{\lfloor#1\rfloor}}\newcommand\ceil[1]{{\lceil#1\rceil}}

\newcommand{\lea}{<^+}
\newcommand{\gea}{>^+}
\newcommand{\eqa}{=^+}

\newcommand{\lel}{<^{\log}}
\newcommand{\gel}{>^{\log}}
\newcommand{\eql}{=^{\log}}

\newcommand{\lem}{\stackrel{\ast}{<}}
\newcommand{\gem}{\stackrel{\ast}{>}}
\newcommand{\eqm}{\stackrel{\ast}{=}}

\newcommand\edf{{\,\stackrel{\mbox{\tiny def}}=\,}}
\newcommand\edl{{\,\stackrel{\mbox{\tiny def}}\leq\,}}
\newcommand\then{\Rightarrow}

\newcommand\km{{\mathbf {km}}}\renewcommand\t{{\mathbf {t}}}
\newcommand\KM{{\mathbf {KM}}}\newcommand\m{{\mathbf {m}}}
\newcommand\md{{\mathbf {m}_{\mathbf{d}}}}\newcommand\mT{{\mathbf {m}_{\mathbf{T}}}}
\newcommand\K{{\mathbf K}} \newcommand\I{{\mathbf I}}

\newcommand\II{\hat{\mathbf I}}
\newcommand\Kd{{\mathbf{Kd}}} \newcommand\KT{{\mathbf{KT}}} 
\renewcommand\d{{\mathbf d}} 
\newcommand\D{{\mathbf D}}

\newcommand\w{{\mathbf w}}
\newcommand\Ks{\Lambda} \newcommand\q{{\mathbf q}}
\newcommand\E{{\mathbf E}} \newcommand\St{{\mathbf S}}
\newcommand\M{{\mathbf M}}\newcommand\Q{{\mathbf Q}}
\newcommand\ch{{\mathcal H}} \renewcommand\l{\tau}
\newcommand\tb{{\mathbf t}} \renewcommand\L{{\mathbf L}}
\newcommand\bb{{\mathbf {bb}}}\newcommand\Km{{\mathbf {Km}}}
\renewcommand\q{{\mathbf q}}\newcommand\J{{\mathbf J}}
\newcommand\z{\mathbf{z}}

\newcommand\B{\mathbf{bb}}\newcommand\f{\mathbf{f}}
\newcommand\hd{\mathbf{0'}} \newcommand\T{{\mathbf T}}
\newcommand\R{\mathbb{R}}\renewcommand\Q{\mathbb{Q}}
\newcommand\N{\mathbb{N}}\newcommand\BT{\Sigma}
\newcommand\FS{\BT^*}\newcommand\IS{\BT^\infty}
\newcommand\FIS{\BT^{*\infty}}\newcommand\C{\mathcal{L}}
\renewcommand\S{\mathcal{C}}\newcommand\ST{\mathcal{S}}
\newcommand\UM{\nu_0}\newcommand\EN{\mathcal{W}}

\newcommand{\supp}{\mathrm{Supp}}

\newcommand\lenum{\lbrack\!\lbrack}
\newcommand\renum{\rbrack\!\rbrack}

\renewcommand\qed{\hfill\emm\square}

\title{\vspace*{-3pc} On the Conditional Complexity of Sets of Strings}

\author {Samuel Epstein\footnote{JP Theory Group. samepst@jptheorygroup.org}}

\maketitle

\begin{abstract}
	Given a set $X$ of finite strings, one interesting question to ask is whether there exists a member of $X$ which is simple conditional to all other members of $X$. Conditional simplicity is measured by low conditional Kolmogorov complexity. We prove the affirmative to this question for sets that have low mutual information with the halting sequence. There are two results with respect to this question. One is dependent on the maximum conditional complexity between two elements of $X$, the other is dependent on the maximum expected value of the conditional complexity of a member of $X$ relative each member of $X$. 
	\end{abstract}
\section{Introduction}

In \cite{Romashchenko03}, a criteria for the amount of algorithmic information that can be extracted from a triplet of strings was established. In that paper, the notion of bunches was introduced. A $(k,l,n)$ bunch is a finite set of strings $X$ such that
\begin{enumerate}
	\item $|X|=2^k$,
	\item $\K(x_1|x_2)<l$ for all $x_1, x_2\in X$,
	\item $\K(x)<n$ for all $x\in X$.
\end{enumerate}
The term $\K$ used above represents the conditional Kolmogorov complexity. In \cite{Romashchenko03}, Theorem 5, it was shown that common information could be extracted from bunches.\\

\noindent\textbf{Theorem 5.} \cite{Romashchenko03}
\textit{For any $(k,l,n)$ bunch $X$, there exists a string $z$ such that $\K(x|z)\leq l +O(|l-k|+\log n)$ and $\K(z|x)=O(|l-k|+\log n)$  for any $x\in X$.}\\
	
In our paper, we revisit bunches and show that every bunch that is not exotic has an element that is simple conditional to all other members. We show this over the class of non-exotic bunches, that is bunches whose encoding has low mutual information with the halting sequence.
We also prove a similar result for a structure we call batches, which are defined in terms of expectation instead of max. In this paper, we use a slightly different definition of bunches (and batches), where there are no assumptions about the Kolmogorov complexity of its elements. We define a $(k,l)$ bunch $X$ to be a finite set of strings, where $k =\ceil{\log |X|}$, $l>k$, and for all $x,x'\in X$, $\K(x|x')\leq l$. If $l\gg k$, such as the bunch consisting of two large independent random strings, then it is difficult to proof properties about it. If $l\approx k$, then interesting properties emerge, such as the bunch theorem of this paper. This theorem states when $l\approx k$, then for non-exotic bunches, there exists common information in the form of a member of this bunch which is simple relative to all other strings of the bunch. Otherwise the bunch is exotic, in that it has high mutual information with the halting sequence. The bunch theorem of this paper is as follows.\\

\noindent\textbf{Theorem.} 
\textit{For $(k,l)$ bunch $X$, $\min_{x\in X}\max_{x'\in X}\K(x|x')\lel 2(l-k)+\I(X:\ch)$.}\\

We also prove a similar result using expectation instead maximum. We define a $(k,l)$ batch $X$ to be a finite set of strings, where $k =\ceil{\log |X|}$, $l>k$, and for all $x\in X$, $\E_{x'\in X}[\K(x|x')]\leq l$.  \\

\noindent\textbf{Theorem.} 
\textit{For $(k,l)$ batch $X$, $\min_{x\in X}\E_{x'\in X}[\K(x|x')]\lel l-k +\I(X:\ch)$.}\\

An example of an exotic bunch is $R_n$, the set of all random strings of size $n$, where $x\in R_n$ iff $\|x\|=n$ and $\K(x)\gea n$. It is not hard to see that for all $x,x'\in R_n$, $\K(x|x')\lel n$. So $R_n$ is a $(n-O(1),n+O(\log n))$ bunch. In addition, because $R_n$ contains all random strings of size $n$, $\min_{x\in X}\max_{x'\in X}\K(x|x')\gel n$. Thus $R_n$ does not have such a conditionally simple element, and this implies it is exotic, because, due to the bunch theorem introduced above, $n\lel \I(R_n:\ch)$. This bound is easily verifiable using the definition of $R_n$, since $\K(R_n)\gea n$ and $\K(R_n|\ch)\lea\K(n)$, because given the halting sequence and $n$, there exists a simple program that can produce all random strings of size $n$.

Another example of a bunch is the set $S_{x,m}$, where $x$ is a string of arbitrary length, and $S_{x,m} =\{xy:y\textrm{ is a string of length }m\}$. This bunch is usually not exotic. It must be that for $\max_{y,x'\in S_{x,m}}\K(y|x')\lea m+\K(m)$ as all strings in $S_{x,m}$ differ by a substring of size $m$. Furthermore $\# S_{x,m}=m$. Therefore $S_{x,m}$ is a $(m,m+\K(m)+O(1))$ bunch. Since $x$ and $m$ can be recovered from an encoding of the the set $S_{x,m}$, and of course $S_{x,m}$ can be created from $x$ and $m$, we have that $\I(S_{x,m}:\ch)\eqa\I(x,m:\ch)<\I(x:\ch)+O(\K(m))$. So by the above bunch theorem, $\min_{y\in S_{x.m}}\max_{x'\in S_{x,m}}\K(y|x')\lel 2\K(m)+\I(S_{x,m}:\ch)\lel \I(x:\ch)+O(\K(m))$. Most $x$ has negligible information with the halting sequence, relative to its length. Furthermore it can be seen independently that $\min_{y\in S_{x,m}}\max_{x'\in S_{x,m}}\K(y|x')\lea\K(m)$, because for $y=x0^m\in S_{x,m}$, there is a program that given any member of $S_{x,m}$ and a program for $m$, can output $y$.
\section{Related Work}

The study of Kolmogorov complexity originated from the work of~\cite{Kolmogorov65}. The canonical self-delimiting form of Kolmogorov complexity was introduced in~\cite{ZvonkinLe70} and treated later in~\cite{Chaitin75}. The universal probability $\m$ was introduced in~\cite{Solomonoff64}. More information about the history of the concepts used in this paper can be found the textbook~\cite{LiVi08}. 

The two main results of this paper, involving bunches and batches, are inequalities including the mutual information of the encoding of a finite set with the halting sequence. A history of the origin of the mutual information of a string with the halting sequence can be found in~\cite{VereshchaginVi04v2}.

A string is stochastic if it is typical of a simple elementary probability distribution. A string is typical of a probability measure if it has a low deficiency of randomness. In the proofs of Theorems \ref{thr:batch} and \ref{thr:bunches}, the stochasticity measure of encodings of finite sets is used. The notion of the deficiency of randomness with respect to a measure follows from the work of~\cite{Shen83}, and also studied in~\cite{KolmogorovUs87,Vyugin87,Shen99}. Aspects involving stochastic objects were studied in~\cite{Shen83,Shen99,Vyugin87,Vyugin99}. 

This work uses the notion of left total machine and the notion of the infinite ``border'' sequence, which is equal to the binary expansion of Chaitin's Omega, (see Section \ref{sec:LeftTotal}). The works of ~\cite{VereshchaginVi04v2,GacsTrVi01} introduced the notion of using the prefix of the border sequence to define strings into a two part code.

This paper can be seen as an update to main result in \cite{EpsteinLe11}, focusing on conditional complexity instead of algorithmic probability. An accessible game-theoretic proof to \cite{EpsteinLe11} can be found in \cite{Shen12}. This paper uses theorems and lemmas found \cite{Epstein13}. Bunches were first introduced by ~\cite{Romashchenko03}, who used them to prove properties of common information of strings.
\section{Conventions}
\label{sec:conv}
We use $\N$, $\Q$, $\R$, $\BT$, $\FS$, and $\IS$ to represent natural numbers, rational numbers, reals, bits, finite strings, and infinite strings. Let $X_{\geq 0}$ and $X_{>0}$ be the sets of non-negative and of positive elements of $X$. The indicator function of a mathematical statement $A$ is denoted by $[A]$, where if $A$ is true then $[A]=1$, otherwise $[A]=0$.

\subsection{Strings}
The length of a string $x{\in}\BT^n$ is denoted by $\|x\|=n$. The removal of the last bit of a string is denoted by $(p0^-){=}(p1^-){=}p$, for $p\in\FS$. For the empty string $\emptyset$, $(\emptyset^-)$ is undefined. We use $\FIS$ to denote $\FS{\cup}\IS$, the set of finite and infinite strings.  For $x\in \FIS$, $y\in \FIS$, we say $x\sqsubseteq y$ iff $x=y$ or $x\in\FS$ and $y=xz$ for some $z\in \FIS$. We say $x\sqsubset y$ if $x\sqsubseteq y$ and $x\neq y$. The $i$th bit of a string $x\in\FIS$ is denoted by $x[i]$. The first $n$ bits of a string $x\in\FIS$ is denoted by $x[0..n]$. 

\subsection{Sets}
The size of a finite set $S$ is denoted to be $|S|$ and also $\# S$ = $\ceil{\log |S|}$. For a finite set $S\subset\FS$, and function $f:\FS\rightarrow\R$, $\E_{x\in S}[f(x)]=\frac{1}{|S|}\sum_{x\in S}f(x)$. 

\subsection{Big O Notation}
As is typical of the field of algorithmic information theory, the theorems in this paper are relative to a fixed universal  machine, and therefore their statements are only relative up to additive and logarithmic precision. For positive real functions $f$ the terms  ${\lea}f$, ${\gea}f$, ${\eqa}f$ represent ${<}f{+}O(1)$, ${>}f{-}O(1)$, and ${=}f{\pm}O(1)$, respectively. In addition ${\lem}f$, ${\gem}f$, and ${\eqm}$ denote $<f/O(1)$, $>f/O(1)$ and $=f*/O(1)$, respectively. For nonnegative real function $f$, the terms ${\lel}f$, ${\gel} f$, ${\eql}f$ represent the terms ${<}f{+}O(\log(f{+}1))$, ${>}f{-}O(\log(f{+}1))$, and ${=}f{\pm}O(\log(f{+}1))$, respectively. 
\subsection{Measures}
A discrete measure is a nonnegative function $Q:\N\rightarrow \R_{\geq 0}$ over natural numbers. The support of a measure $Q$ is the set of all elements $a\in\N$ that have positive measure, with $\supp(Q) = \{a\,{:}\,Q(a)>0\}$. 
\begin{dff}[Elementary Measures]
	\label{def:Elem}
The measure is elementary if its support is finite and its range is a subset of $\Q$. 
\end{dff}
Elementary measures have an explicit finite encoding, in the natural way. The mean of a function $f:\N\rightarrow\R$ by a measure $Q$ is denoted by $\mathbf{E}_{a\sim Q}[f] = \sum_{a\in\N} f(a)Q(a)$. We say $Q$ is a semimeasure iff $\mathbf{E}_{a\sim Q}[1]\,{\leq}\,1$. Furthermore, we say that $Q$ is probability measure iff $\mathbf{E}_{a\sim Q}[1]\,{=}\,1$.  For a set $S\subseteq\N$, $Q(S)=\sum_{x\in S}Q(x)$. For semimeasure $Q$, we say that $d:\N\rightarrow\R_{\geq 0}$ is a $Q$ test, if $\E_{a\sim Q}[2^{d(a)}]\leq 1$.

\subsection{Algorithms and Complexity}
$T_y(x)$ is the output of algorithm $T$ (or $\perp$ if it does not halt) on input $x\in\FS$ and auxiliary input $y\in\FIS$. $T$ is prefix-free if for all $x,s\in\FS$ with $s\,{\neq}\,\emptyset$, and $y\in\FIS$, either $T_y(x)\,{=}\perp$ or $T_y(xs)\,{=}\perp$ . The complexity of $x\in\FS$ with respect to $T_y$ is $\K_T(x|y)= \inf\{\|p\|\,:\,T_y(p)=x\}$. 

There exist optimal for $\K$ prefix-free algorithm $U$, meaning that for all prefix-free algorithms $T$,  there exists $c_T\,{\in}\,\N$, where $\K_U(x|y)\leq \K_T(x|y)+c_T$ for all $x\,{\in}\,\FS$ and $y\,{\in}\,\FIS$. For example, one can take a universal prefix-free algorithm $U$, where for each prefix-free algorithm $T$, there exists  $t\in\FS$, with $U_y(tx)=T_y(x)$ for all $x\in\FS$ and $y\in\FIS$. $\K(x|y)$ is defined to be $\K_U(x|y)$ is the Kolmogorov complexity of $x\in\FS$ relative to $y\in\FIS$. When we say that universal Turing machine is relativized to an object, this means that an encoding of the object is provided to the universal Turing machine on an auxilliary tape. A function $f:\N\rightarrow\N$ is partial computable with respect to $U$ if there is a string $t\in\FS$ such that $f(x)=U(t\langle x\rangle)$ when $f(x)$ is defined and $U(t\langle x\rangle)$ does not halt otherwise.

\begin{dff}[Complexity of Computable Function]
	\label{def:ComplComp}
The complexity of a (partial) computable function $f:\N\rightarrow\N$, is $\min_{i\in D_f}\K(i)$ where $D_f$ is the set of indices of functions equal to $f$ in an enumeration of partial computable functions of the form $\N\rightarrow\N$. 
\end{dff}
A function $f:\FS\rightarrow\R_{\geq 0}$ is lower semicomputable if the set $\{(x,q)\,{:}\,x\in\FS, f(x)>q\in\Q\}$ is enumerable.
\begin{dff}[Complexity of Semicomputable Function] 
\label{def:ComplLowerComp}	
The complexity of a lower semicomputable function $f$ is $\min_{i\in G_f}\K(i)$, where $G_f$ is the set of indices of functions that enumerate $\{(x,q),x\in\FS, f(x)>q\in\Q\}$ in an enumeration of all enumerations that outputs a subset of $\FS\times\Q$.
\end{dff}
\subsection{Properties of Complexity}
The chain rule for Kolmogorov complexity is $\K(x,y) \eqa \K(x)+\K(y|\langle x,\K(x)\rangle)$.  The universal probability of a set $D\subseteq\FS$ is $\m(D|y){= }\sum_z[\,U_y(z)\in D]2^{-\|z\|}$. For strings $x\in\FS$, we have $\m(x|y)=\m(\{x\}|y)$. The coding theorem states $-\log \m(x|y)\eqa\K(x|y)$.

The halting sequence $\mathcal{H}\in\IS$ is the infinite string where $\mathcal{H}[i]=[U(i)\neq \perp]$ for all $i\in\N$. We recall that the amount of mutual information that $a\in\N$ has with $\mathcal{H}$ conditional to $b\in\N$ is denoted by $\I(a:\mathcal{H}|b)=\K(a|b)-\K(a|b,\mathcal{H})$.
\section{Left-Total Machines}

\label{sec:LeftTotal}
\begin{figure}[h!]
	\begin{center}
		\includegraphics[width=0.5\columnwidth]{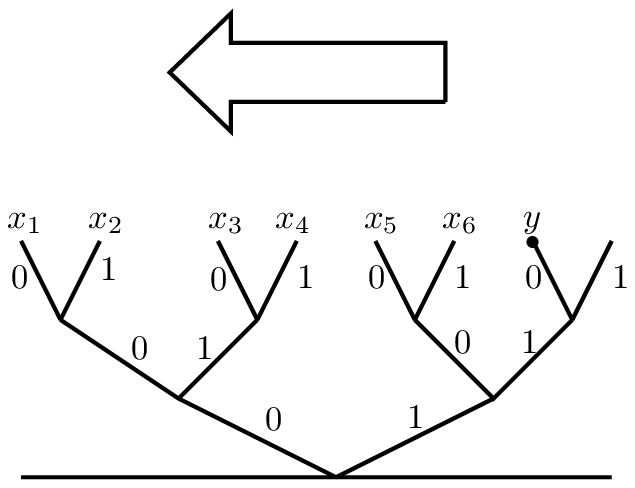}
		\caption{The above diagram represents the domain of a left total machine $T$ with the 0 bits branching to the left and the 1 bits branching to the right. For $i\in \{1,\dots,5\}$, $x_i\lhd x_{i+1}$ and $x_i\lhd y$. Assuming $T(y)$ halts, each $x_i$ is total. This also implies each $x_i^-$ is total as well.}
		\label{fig:LeftTotal}
	\end{center}
\end{figure}

The notions of total strings and the ``left-total'' universal algorithm are needed in the remaining sections of the paper. We say $x\in\FS$ is total with respect to a machine if the machine halts on all sufficiently long extensions of $x$. More formally, $x$ is total with respect to $T_y$ for some $y\in\FIS$ iff there exists a finite prefix free set of strings $Z\subset\FS$ where $\sum_{z\in Z}2^{-\|z\|}=1$ and $T_y(xz)\neq\perp$ for all $z\in Z$.  We say (finite or infinite) string $\alpha\in\FIS$ is to the ``left'' of $\beta\in\FIS$, and use the notation $\alpha\lhd \beta$, if there exists a $x\in\FS$ such that $x0\,{\sqsubseteq}\, \alpha$ and $x1\,{\sqsubseteq}\, \beta$. A machine $T$ is left-total if for all auxiliary strings $\alpha\in\FIS$ and for all $x,y\in\FS$ with $x\lhd y$, one has that $T_\alpha(y)\neq\perp$ implies that $x$ is total with respect to $T_\alpha$. An example can be seen in Figure \ref{fig:LeftTotal}.

For the remaining of this paper, we can and will change the universal self delimiting machine $U$ into a universal left-total machine $U'$ by the following definition. The algorithm $U'$ orders all strings $p\,{\in}\,\FS$ by the running time of $U$ when given $p$ as an input. Then $U'$ assigns each $p$ an interval $i_p{\subseteq}[0,1]$ of width $2^{-\|p\|}$. The intervals are assigned ``left to right'', where if $p\in\FS$ and $q\in\FS$ are the first and second strings in the ordering, then they will be assigned the intervals $[0,2^{-\|p\|}]$ and $[2^{-\|p\|},2^{-\|p\|}\,{+}\,2^{-\|q\|}]$. 

Let the target value of $p\in\FS$ be $(p)\,{\in}\,\mathbb{W}$, which is the value of the string in binary. For example, the target value of both strings 011 and 0011 is 3. The target value of 0100 is 4. The target interval of $p\in\FS$ is $\Gamma(p)=((p)2^{-\|p\|},((p){+}1)2^{-\|p\|})$.

The universal machine $U'$ outputs $U(p)$ on input $p'$ if $\Gamma(p')$ and not that of $\Gamma({p'}^{-})$ is strictly contained in $i_p$. The same definition applies for the machines $U'_\alpha$ and $U_\alpha$, over all $\alpha\,{\in}\,\FIS$. 

Recall that a function $f:\N\rightarrow\N$ is partial computable with respect to $U$ if there is a string $t\in\FS$ such that $f(x)=U(t\langle x\rangle)$ when $f(x)$ is defined and $U(t\langle x\rangle)$ does not halt otherwise. Similarly a function $f:\N\rightarrow\N$ is partial computable with respect to $U'$ if there is $t\in\FS$, that whenever $f(x)$ is defined, there is an interval $i_{t\langle x\rangle}$ and for any string $p$ where $\Gamma(p)$ and not that of $\Gamma(p^-)$, is contained in $i_{t\langle x\rangle}$, then $U'(p)=f(x)$. Otherwise, if $f(x)$ is not defined, there does not exist the interval $i_{t\langle x\rangle}$.

\begin{prp}
	$\K_U(x|y)\eqa\K_{U'}(x|y)$.
	\label{prp:kequiv}
\end{prp}
\begin{prf}
	It must be that $\K_U(x|y)\lea \K_U'(x|y)$, because there is a Turing machine that computes $U'$. Therefore, do to the universality of $U$, there is a $t\in\FS$, such that $U_y(tx)=U'_y(x)$, thus proving the minimality of $\K_U$. It must be that $\K_U'(x|y)\lea \K_U(x|y)$. This is because if $U(x)=z$, then there is interval $i_x$ such that for all strings  $p$ where $\Gamma(p)$ and not that of $\Gamma(p^-)$ that are strictly contained in $i_x$ has $U'_y(p)= U_y(x)$. Thus we have that $\|p\|\leq \|x\|+2$. This implies that $\K_{U'}(x|y)\leq \K_U(x|y)+2$.\qed
\end{prf}\\

For the rest of the paper, we now set $U$ to be equal $U'$, so the universal Turing machine can be considered to be left-total. Without loss of generality, as shown in Proposition \ref{prp:kequiv} the complexity terms of this paper are defined with respect to the universal left total machine $U$.

\begin{prp}
	There exists a unique infinite sequence $\mathcal{B}$ with the following properties. 
	\begin{enumerate}
	\item All the finite prefixes of $\mathcal{B}$ have total and non-total extensions. 
	\item If a finite string has total and non-total extensions then it is a prefix of $\mathcal{B}$.
	\end{enumerate}
\end{prp}
\begin{prf}
	(1) Let $\Omega\in\R$ be the Chaitin's Omega, the probability that a random sequence of bits halts when given to $U$, with $\Omega =\sum_{p\in\FS}[U(p)\textrm{ halts}]2^{-\|p\|}$.  Thus $\Omega$ characterizes the domain of $U'$, with $\bigcup_{p\in\FS}i_p=[0,\Omega)$. Let $\mathcal{B}\in \IS$ be the binary expansion of $\Omega$, which is a ML random string. For each  $n\in \N$, let $b_n\sqsubset\mathcal{B}$, $\|b_n\|=n$. Let $m\in\mathbb{W}$ be the smallest whole number such that $b_n1^{(m)}0\sqsubset\mathcal{B}$. Then $b_n1^{(m+1)}$ is a non-total string because $[0,\Omega]\cap\Gamma(b_n1^{(m+1)})=\emptyset$. Furthermore let $m\in\mathbb{W}$ be the smallest whole number such that $b_n0^{(m)}1\sqsubset\mathcal{B}$. Then $b_n0^{(m+1)}$ is a total string because $\Gamma(b_n0^{(m+1)})\subset[0,\Omega)$.
	
	(2) Assume there are two strings $x$ and $y$ of length $n$ that have total and non-total extensions, with $x\lhd y$. Since $y$ has total extensions, there exist $z$ such that $U'(yz)$ halts. Since $x\lhd yz$, by the definition of left-total machines, $x$ is total, causing a contradiction.\qed
\end{prf}
\begin{figure}[h!]
	\begin{center}
		\includegraphics[width=0.4\columnwidth]{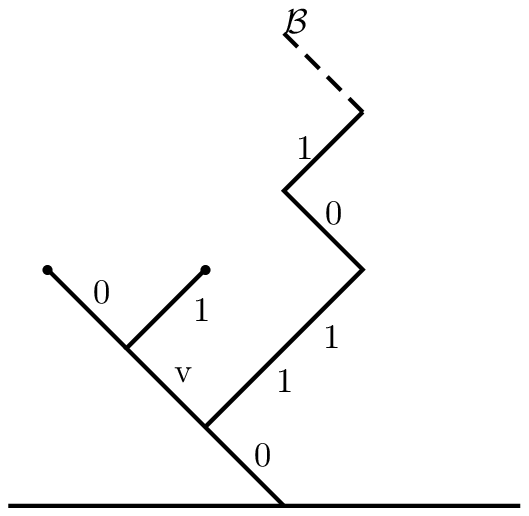}
		\caption{The above diagram represents the domain of the universal left-total algorithm $U'$, with the 0 bits branching to the left and the 1 bits branching to the right. The strings in the above diagram, $0v0$ and $0v1$, are halting inputs to $U'$ with $U(0v0)\neq \perp$ and $U(0v1)\neq \perp$. So $0v$ is a total string. The infinite border sequence $\mathcal{B}\in\IS$ represents the unique infinite sequence such that all its finite prefixes have total and non total extensions. All finite strings branching to the right of $\mathcal{B}$ will cause $U'$ to diverge.}
		\label{fig:DomainUPrime}
	\end{center}
\end{figure}

We call this infinite sequence $\mathcal{B}$, ``border'' because for any string $x\in\FS$, $x\lhd\mathcal{B}$ implies that $x$ is total with respect to $U$ and $\mathcal{B}\lhd x$ implies that $U$ will never halt when given $x$ as an initial input. Figure~\ref{fig:DomainUPrime} shows the domain of $U'$ with respect to $\mathcal{B}$. 

We now set $U$ to be equal $U'$. Without loss of generality, as shown in Proposition \ref{prp:kequiv} the complexity terms of this paper are defined with respect to the universal left total machine $U$.

For total string $b$, let $\mathbf{bbtime}(b)$, be the slowest running time of a program that extends $b$ or is to the left of $b$.  With respect to the universal Turing machine $U$ defined above, $\mathbf{bbtime}(b)$ would be the running time of the rightmost extension of $b$ that halts. For total string $b\in\FS$, and $x,y\in\FS$, let $\m[b](x|y)$ be the algorithmic weight of $x$ from programs conditioned on $y$ in time $\mathbf{bbtime}(b)$. More formally, $\m[b](x|y) = \sum \{2^{-\|p\|}:U_y(p)=x\textrm{ in time }\mathbf{bbtime}(b)\}.$ The term $\m[b](x|y)$ is 0 if $b$ is not total. For every total $b$, $\m[b](\cdot|y)$ is a semi-measure and thus there exists a prefix free codebook that assigns to every string in the support of $\m[b](\cdot|y)$ a codeword of length $\K[b](x|y) = \ceil{-\log \m[b](x|y)}$. Furthermore $\K[b](x|y)$ is defined to be $\infty$ if $\m[b](x|y)$ is 0.

\section{Stochasticity}
\label{sec:stoch}
In algorithmic statistics, a string is stochastic if it is typical of a simple probability measure. Properties of stochastic (and non-stochastic) strings can be found in the survey \cite{VereshchaginSh17}. The deficiency of randomness of $x$ with respect to \textit{elementary} probability measure $Q$ and $v\in \FS$ is $\d(x|Q,v)=\ceil{-\log Q(x)}-\K(x|v)$. Elementary measures are introduced in Definition \ref{def:Elem}. The function $\d(\cdot|Q,v)$ is a $Q$-test (up to an additive constant). It is also universal, in that for any lower semicomputable test $d$, and $v\in\FS$, for all $x\in\FS$, $d(x|Q,v)\lea \d(x|Q,v)+\K(d|v)+\K(Q|v)$, as shown in \cite{Gacs13}. The complexity of $d$ given $v$, $\K(d|v)$ is that of Definition \ref{def:ComplLowerComp}.

For some $j,k\in\N$, we say that $x\in\N$ is $(j,k)$-stochastic if there exists $v\in\BT^j$, with $U(v)=Q$, $Q$ being an elementary probability measure, and $\d(x|Q,v)\leq k$. The stochasticity of $x\in\N$, is measured by $\Ks(x)=\min\{j+3 k\,{:}\,x\textrm{ is $(j,k)$ stochastic}\}$. The conditional stochasticity form\footnote{This is formally represented as $\Ks(x|\alpha) = \min\{j+3k\,{:}\,\exists v\in\{0,1\}^j, U_\alpha(v)=\langle Q\rangle, \d(x|Q, \langle v,\alpha\rangle)\leq k\in\N\}$.}
is represented by $\Ks(x|\alpha)$, for $\alpha\,{\in}\,\FIS$. 

Stochasticity follows non-growth laws; a total computable function cannot increase the stochasticity of a string by more than a constant factor dependent on its complexity. Lemma \ref{lem:StochTotalF} illustrates this point. The complexity of total function $f$, $\K(f)$, is that of Definition \ref{def:ComplComp}. Another variant of the same idea can be found in Proposition 5 in \cite{VereshchaginSh17}.

 \begin{lmm}
 	\label{lem:StochTotalF}
 	Given total computable function $f:\FS\rightarrow\FS$, $\Ks(f(x))< \Ks(x)+O(\K(f))$.
	\end{lmm}
\begin{prf}
	Let $v\in\FS$ realize $\Ks(x)$, with $U(v)=Q$, $\|v\|+3\max\{\d(x|Q,v),1\}=\Ks(x)$.  Let $f(Q)$ be the image distribution of $Q$ with respect to $f$. Thus $f(Q)(a)=\sum_{b:f(b)=a}Q(b)$. The function $\d(f(\cdot)|f(Q),v)$ is a $Q$ test (relative to $v$ and up to an additive constant), because $$\sum_{a\in\FS}2^{\d(f(a)|f(Q),v)}Q(a)=\sum_{b\in\FS}2^{\d(b|f(Q),v)}f(Q(b))< O(1).$$ 
Also $\d(f(\cdot)|f(Q),v)$ is lower semi-computable given $v$, with $\K(\d(f(\cdot)|f(Q),v)|v)\lea\K(f|v)$. So due to the universality of $\d$, $\d(f(x)|f(Q),v)\lea \d(x|Q,v)+\K(f|v)\lea \d(x|Q,v)+\K(f).$ Let $v' = v_0vv_f\in\FS$ compute $f(Q)$, where $v_0$ is helper code of size $O(1)$ and $v_f$ is a shortest program that computes $f$, with $\|v_f\|=\K(f)$ . So $\|v'\| \lea \|v\| +\K(f)$.  Since $\K(x|v)\lea \K(x|v')+\K(v'|v)\lea \K(x|v')+\K(f)$, we have that $\d(f(x)|f(Q),v')\lea \d(x|Q,v)+O(\K(f))$. So
	\begin{align*}
	\Ks(f(x)) &\leq \|v'\|+3\max\{ \d(f(x)|f(Q),v'),1\}\\
	&\lea \|v\|+3\max\{ \d(f(x)|f(Q),v'),1\}+\K(f)\\
	&<\|v\|+3\max\{\d(x|Q,v),1\}+O(\K(f))\\
	&\leq\Ks(x)+O(\K(f)).
	\end{align*}\qed
\end{prf}
\newpage

The following lemma is taken from \cite{EpsteinLe11}. It states that the stochasticity measure of a string lower bounds its information with the halting sequence. 

\begin{lmm}
	\label{lem:StochH}
	For $x\in\FS$, $\Ks(x)\lel\I(x:\ch)$.
\end{lmm}

\begin{prf}
	Let $U(x^*)=x$, $\|x^*\|=\K(x)$, and $v$ be the shortest total prefix of $x^*$. We define the elementary probability measure $Q$ such that $Q(a) = \sum_w 2^{-\|w\|}[U(vw)\,{=}\,a]$. Thus $Q$ is computable relative to $v$. In addition, since $v\sqsubseteq x^*$, one has the lower bound $Q(x) \geq  2^{-\|x^*\|+\|v\|} = 2^{-\K(x)+\|v\|}$. Therefore 
	\begin{align}
	\nonumber
	\mathbf{d}(x|Q,v) &= \ceil{-\log Q(x)} - \K(x|v) \\
	\nonumber
	&\lea  \K(x)-\|v\| - \K(x|v)\\ 
	\nonumber
	&\lea (\K(v) +\K(x|v)) - \|v\| -\K(x|v)\\
	\nonumber
	&\lea (\|v\|+\K(\|v\|)+\K(x|v)) - \|v\| -\K(x|v),\\
	\mathbf{d}(x|Q,v) &\lea \K(\|v\|).\label{eqnDef}
	\end{align}
	
	Since $v$ is total and $v^{-}$ is not total, by  Proposition~\ref{prp:borderprefix}, $v^-$ is a prefix of the border sequence $\mathcal{B}$. In addition, $Q$ is computable from $v$. Therefore
	\begin{align}
	\nonumber
	\K(x|\ch) &\lea \K(x|Q) +\K(Q|\ch)\\
	\nonumber 
	&\lea \K(x|Q) + \K(\ch)\\
	&\lea -\log Q(x)+\K(\|v\|)\label{eq:Halting1}\\ 
	\nonumber
	&\lea \K(x) - \|v\| +\K(\|v\|),\\
	\nonumber
	\|v\| &\lea \K(x)-\K(x|\ch) +\K(\|v\|),\\
	\|v\| &\lel \I(x;\ch).\label{eq:Halting2}
	\end{align}
	
	Equation~(\ref{eq:Halting1}) is due $\mathcal{B}$ being computable from $\mathcal{H}$, therefore $v^-\sqsubset\mathcal{B}$ is simple relative to  $\mathcal{H}$ and $\|v\|$. Since $Q$ is computable from $v$, one gets $\Ks(x) \lea \K(v)+3\log(\max\{\mathbf{d}(x|Q,v),1\})\lea \|v\|+\K(\|v\|)+3\log(\max\{\mathbf{d}(x|Q,v),1\})$. Due to equation~\ref{eqnDef}, one gets $\Ks(x) \leq \|v\| + O( \K(\|v\|))\lel \|v\|$. Due to equation~\ref{eq:Halting2}, one gets $\Ks(x)\lel \I(x;\ch)$.\qed
\end{prf}\\

The following lemma shows that if a prefix of the border sequence is simple relative to a string $x$, then it will be the common information between $x$ and the halting sequence $\ch$. Note that if a string $b$ is total and $b^-$ is not, then $b^-\sqsubset\mathcal{B}$, due to the fact that $b^-$ has total and non-total extensions.

\begin{prp}
	\label{prp:borderprefix}
	The border sequence $\mathcal{B}$ is Martin L\"{o}f random, where for $b\sqsubset\mathcal{B}$, $\|b\|\lea \K(b)$. Furthermore if $b\in\FS$ is total and $b^-$ is not, then $b^-\sqsubset\mathcal{B}$.
\end{prp}
\begin{prf}
	Let $\Omega =\sum_x\m(x)$ be Chaitin's Omega, the probability that U will halt. It is well known that the binary expansion $\Omega'\in\IS$ of $\Omega$ is Martin L\"{o}f random. Given $b\sqsubset\mathcal{B}$, $b\in\BT^n$, one can compute $\hat{\Omega}=\sum\{2^{-\|y\|}[U(y)\neq\perp] : y\lhd b\}$ with differs from $\Omega$ in the summation of programs which branch from $\mathcal{B}$ at positions $n$ or higher. Thus $\Omega-\hat{\Omega}\leq 2^{-n}$. So $n\lea \K(\Omega'[0..n-1]) \lea \K(\Omega'[0..n-1],b)\lea \K(\Omega'[0..n-1]|b)+\K(b)\lea \K(b)$. Thus $\mathcal{B}$ is Martin L\"{o}f Random. If $b\in \FS$ is total and $b^-$ is not, then $b^-$ has a total extension $b^-0$ and a non total extension $b^-1$, thus by the definition of the border sequence, $b^-\sqsubset\mathcal{B}$.\qed\\
\end{prf}\newpage
\begin{lmm}
	\label{lmm:totalString}
	If $b\in\FS$ is total and $b^-$ is not, and $x\in\FS$, 
	then $\K(b)+\I(x;\mathcal{H}|b)\lel \I(x\,{;}\,\mathcal{H})+\K(b|x)$.
\end{lmm}

\begin{prf}
	By Proposition \ref{prp:borderprefix}, $b^-\sqsubset\mathcal{B}$ is a prefix of the border sequence and thus $\|b\|\lea\K(b)$. Since $\mathcal{B}$ is computable from the halting sequence $\mathcal{H}$, we have that $b$ is computable from $\|b\|$ and $\mathcal{H}$, with $\K(b|\mathcal{H})\lea \K(\|b\|)$. The chain rule gives the equality $\K(b)+\K(x|b,\K(b)) \eqa \K(x)+\K(b|x,\K(x))$. Combined with the inequalities $\K(x|b)\lea \K(x|b,\K(b))+\K(\K(b))$ and $\K(b|x,\K(x))\lea \K(b|x)$, we get
	\begin{align*}
	\K(b)+\K(x|b)&\lea \K(x)+\K(b|x)+\K(\K(b)).
	\end{align*} 
	
	Subtracting $\K(x|b,\mathcal{H})$ from both sides results in
	\begin{align*}
	\K(b)+\K(x|b)-\K(x| b,\mathcal{H})&\lea \K(x)+\K(b|x)+\K(\K(b))-\K(x| b,\mathcal{H})\\
	&\lea \K(x)+\K(b|x)+\K(\K(b))-\K(x|\mathcal{H})+\K(b|\mathcal{H}).\\
	&\lea\I(x;\mathcal{H})+\K(b|x)+\K(\K(b))+\K(b|\mathcal{H})\\
	&< \I(x;\mathcal{H})+\K(b|x)+O(\log\|b\|)\\
	&\lel\I(x;\mathcal{H})+ \K(b|x).
	\end{align*}\qed
\end{prf}$ $\\

The following theorem is from \cite{EpsteinLe11}.  Another proof of this theorem can be found in \cite{Shen12}. It states that sets that are not exotic, i.e. sets with low mutual information with the halting sequence, have simple members that contain a large portion of the algorithmic weight of the sets. It is compatible with this paper's stochasticity definition because the term $\Ks$ used in this paper is larger than the stochasticity measure used in \cite{EpsteinLe11}. 

\begin{thr}
	\label{thr:el} For finite set $D\subset\FS$, $\min_{x\in D}\K(x)\lea \ceil{-\log\m(D)}+2\K(\ceil{-\log\m(D)})+\Ks(D)$.
\end{thr}
\section{Batches}
\label{sec:batches}
Recall that a $(k,l)$ batch $X$ is a finite set of strings, where $k = \#X$, $l>k$, and for all $x\in X$, $\E_{x'\in X}[\K(x|x')]\leq l$. The following theorem states that for non-exotic batches, there is an element of $X$ that is simple, on average, conditional to all other members of $X$.
\begin{thr}
	\label{thr:batch}
For $(k,l)$ batch $X$, $\min_{x\in X}\E_{x'\in X}[\K(x|x')]\lel l-k +\I(X:\ch)$.
\end{thr}

\noindent \textbf{Informal Proof.}\\

\noindent \textit{This proof uses a probability distribution $\kappa$ over functions $g:S\rightarrow\N$, for some finite set $S$ where $\kappa(g)=\prod_{a\in S}2^{-g(a)}$. Thus functions with smaller values have higher $\kappa$ measure. We define a conditional measure $P_g(y|x')\approx \m[b](y|x')2^{g(y)}$. Using probabilistic arguments, we show there exists some $g$ where $P_g(y|x)$ is large enough for all $y\in S$ and $x\in X$ and there is an $x_g\in X$, where $g(x_g)\eqa \# X$. Thus $\E_{x'\in X}[\K(x_g|x')]\lessapprox\E_{x'\in X}[-\log P_g(x_g|x')]\lessapprox \E_{x'\in X}[\K[b](x_g|x')-g(x_g)]\lessapprox \E_{x'\in X}[\K[b](x_g|x')-\#X]\lessapprox l-k$. The terms $\m[b]$ and $\K(b)$ are computable complexity and algorithmic probability using a total string $b$ that is factored out at the end of the proof. In the above inequality, there is an additional $\K(g|b)$ term. Using the fact that there is a large $\kappa$ measure of suitable $g$ candidates, and Theorem \ref{thr:el} and Lemmas \ref{lem:StochTotalF} and \ref{lem:StochH}, we get that $\K(g|b)\lel \I(X:\ch|b)$. Using Lemma \ref{lmm:totalString}, the total string $b$ is removed from the inequality.}

\begin{prf}$ $\\
	
\noindent\textbf{(1).} \textit{The first step of the proof is to find a total string such that $X$ is a batch with computable complexity $\K[b]$, with $\max_{y\in X}\E_{x'\in X}[\K[b](y|x')]\lea l$. This enables the proof to use computable complexity and algorithmic probability $\K[b]$ and $\m[b]$ instead of their semi-computable counterparts. The string $b$ is factored out at the end of the proof in Section 9.} \\

\noindent We can assume that $k>2$, otherwise the theorem is trivially proven. Let $b$ be the shortest total string where $\max_{y\in X}\E_{x'\in X}[\K[b](y|x')]<l+2$, dubbed property $A$. Thus $\K(b|X)\lea\K(\|b\|,(l\,{-}\,k))$. This is because, firstly, $l$ can be constructed from $(l-k)$ and $X$. This is because from $X$, one can compute $\#X$, and thus $k$. Then from a program that computes $(l-k)$ and $k$, one can compute $l$. Secondly there exists a program that can enumerate all total strings of length $\|b\|$ from ``left'' to ``right''. For each enumerated total string $h$ of length $\|b\|$, one can compute $\K[h]$ for all strings, and thus $\max_{y\in X}\E_{x'\in X}[\K[h](y|x')]$. This program can select the first one with property $A$. The first one selected will be $b$, otherwise there exists a $b'\lhd b$, $\|b'\|=\|b\|$, with property $A$. This implies there exists a total ${b'}^-$ with $\K[{b'}^-]\leq\K[b']$. Thus property $A$ holds for ${b'}^-$, contradicting the minimal length of $b$. This also implies $b^-$ is not total. \\

\noindent\textbf{(2).} \textit{We define a probability measure $\kappa$ over functions $g\in\mathcal{G}$ from the support of $\m[b]$ to natural numbers, where functions with low values will have a higher probability.}\\

\noindent  Let $S=\supp(\m[b])$ be the support of $\m[b]$, which is finite. Let $\mathcal{G}$ be the infinite set of all functions $g:S\rightarrow\N$. Since $S$ is finite, each $g\in\mathcal{G}$ can be encoded in an explicit finite string. Let $\kappa:\mathcal{G}\rightarrow\R_{\geq 0}$ be a probabilility measure where $\kappa(g)=\prod_{a\in S}2^{-g(a)}$. So for all $a\in S$, it must be that $\kappa(\{g:g(a)=n\})=2^{-n}$ and $\kappa(\{g:g(a)\geq n\})=2^{-n+1}$.\\  
	
\noindent\textbf{(3).} \textit{The set $\mathcal{G}^H_1\subset\mathcal{G}$ is defined. It is the class of functions $g$ which has a low enough $g(x_g)$ for some $x_g$. Using probabilistic arguments, it is shown that $\mathcal{G}^H_1$ has significant $\kappa$ measure.}\\

\noindent For any finite set $H\subset\FS$, $\#H>2$, let $\mathcal{G}^H_1$ be the set of functions $g\in \mathcal{G}$, where there exists $x_g\in H$ with $g(x_g)=\# H-2$. Using the fact that $(1-m)e^m\leq 1$ for $m\in[0,1]$, we have that $$\kappa(\mathcal{G}\,{\setminus}\,\mathcal{G}^H_1 )\leq\prod_{a\in H}\left(1-2^{-\# H+2}\right)\leq \left(1-2^{-\# H+2}\right)^{2^{\# H-1}}\leq e^{-2^{-\# H+2}2^{\# H-1}}= e^{-2}< 0.25.$$
So $\kappa(\mathcal{G}^H_1)>0.75$.\\

\noindent\textbf{(4).} \textit{We define the conditional measure $P'_g$, indexed by $g\in\mathcal{G}$. Though it contains complicating convergerence and boundary condition terms, its essence is $P'_g(y|x') \approx 2^{g(y)}\m[b](y|x')$. Thus the function $g\in\mathcal{G}$ has property of boosting up values of $\m[b]$. }\\

\noindent We use measures $P'_g(y|x'):\FS\rightarrow\R_{\geq 0}$, indexed by $g\in\mathcal{G}$ and $x'\in S$. The measure $P'$ is defined as $P'_g(y|x')=[\delta_g(y,x')> 0]2^{-\delta_g(y,x')}\delta_g(y,x')^{-2}+[\delta_g(y,x')\leq 0]$, where $\delta_g(y,x')=\K[b](y|x')-g(y)$. By the definition of measurement, for a set $B\subseteq \FS$, we have that $P'_g(B|x')=\sum_{a\in B}P'_g(a|x')$.  \\

\noindent\textbf{(5).}\textit{We introduce a second set of functions $\mathcal{G}^H_2\subset\mathcal{G}$. If $g\in\mathcal{G}^H_2$, then the $P'$-normalizing term $P'_g(S|x')$ is small, over the expectation of $x'\in H$.  Thus if $g\in \mathcal{G}^H_2$, it can be used to create a useful probability measure $P_g$, defined in Section 6 of the proof. Using the Markov inequality, $\mathcal{G}^H_2$ also has significant $\kappa$ measure.}\\
 
\noindent We define a second set of functions $\mathcal{G}^H_2=\{g:\E_{x'\in H}[P_g'(S|x')]\leq 8, g\in\mathcal{G}\}$. The bound of 8 is chosen to satisfy a Markov inequality later in the proof. So
\begin{align*}
&\;\;\;\; \E_{g\sim\kappa}\E_{x'\in H}[P'_g(S|x')]\\
&= |H|^{-1}\sum_{x'\in H,y\in S}\E_{g\sim\kappa}[P'_g(y|x')]\\
&= |H|^{-1}\sum_{x'\in H}\sum_{y\in S}\left(\sum_{c=1}^{\K[b](y|x')-1}2^{c-\K[b](y|x')}(\K[b](y|x')-c)^{-2}\kappa(\{g:g\in\mathcal{G},g(y)=c\})\right)\\
&\hspace{3cm} + \kappa(\{g:g\in\mathcal{G},g(y)\geq \K[b](y|x')\})\\
&\leq |H|^{-1}\sum_{x'\in H}\sum_{y\in S}\left(\m[b](y|x')\sum_{c=1}^{\K[b](y|x')-1}(\K[b](y|x')-c)^{-2}\right)+2^{-\K[b](y|x')+1}\\
&\leq |H|^{-1}\sum_{x'\in H}2\m[b](S|x')+2\m[b](S|x')<4.
	\end{align*}
So by the Markov inequality, $\kappa(\mathcal{G}^H_2)\geq 0.5$. So for all finite $H\subset\FS$, $\# H>2$,  $\kappa(\mathcal{G}^H_1\cap\mathcal{G}^H_2)>0.25$. \\

\noindent\textbf{(6).} \textit{The probability $P_g$ is defined as a normalization of $P'_g$ over $S$. If $g$ is in both $\mathcal{G}_1^H$ and $\mathcal{G}_2^H$, then there is a $x_g\in H$, where the expection of $-\log P_g(x_g|x')$ over $x'\in H$ is less than that of the expection of $-\log P'_g(x_g|x')$, up to an additive constant. Furthermore, since $g\in \mathcal{G}^H_1$, this value is bounded, up to logarithmic precision, to the expectation $\K[b](x_g|x')-\#H$, over all $x'\in H$.}\\
	
\noindent We use the following probability measure $P_g(y|x')$, indexed by $g\in\mathcal{G}$ and $x'\in S$, defined as $P_g(y|x')=[y\in S]P'_g(y|x')/P_g'(S|x')$. Thus $P_g(\FS|x')=1$ for all $x'\in S$, $g\in\mathcal{G}$. So for any $g\in \mathcal{G}^H_1\cap\mathcal{G}^H_2$, there exists $x_g\in H$ where $g(x_g)=\#H -2$ and also
\begin{align}
\nonumber
&\;\;\;\; \E_{x'\in H}[-\log P_g(x_g|x')]\\
\label{eq:PKb01}
&= \E_{x'\in H}[-\log P_g'(x_g|x')+\log P_g'(S|x')]\\
\nonumber
&=\E_{x'\in H}[-\log P_g'(x_g|x')]+\E_{x'\in H}[\log P_g'(S|x')]\\
\nonumber
&\leq\E_{x'\in H}[-\log P_g'(x_g|x')]+\log\E_{x'\in H}[ P_g'(S|x')]\\
\label{eq:PKb02}
&\lea\E_{x'\in H}[-\log P_g'(x_g|x')]\\
\label{eq:PKb03}
&\eqa \E_{x'\in H}\left[[\delta_g(x_g,x')> 0](-\log 2^{-\delta_g(x_g,x')}\delta_g(x_g,x')^{-2})+[\delta_g(x_g,x')\leq 0]\right]\\
\nonumber
&\lea  \E_{x'\in H}[\max\{\delta_g(x_g,x')+2\log \delta_g(x_g,x'),O(1)\}]\\
\nonumber
&\lea \max\{ \E_{x'\in H}[\delta_g(x_g,x')]+2\log\E_{x'\in H}[\delta_g(x_g,x')],O(1)\}\\
\label{eq:PKb04}
&\lea \max\{ \E_{x'\in H}[\K[b](x_g|x')-g(x_g)]+2\log\E_{x'\in H}[\K[b](x_g|x')-g(x_g)],O(1)\}\\
\label{eq:PKb1}
&\lel \max\{\E_{x'\in H}[\K[b](x_g|x')]-\# H,O(1)\}.
\end{align}

Equation \ref{eq:PKb01}, follows from definition of $P_g$. Equation \ref{eq:PKb02} follows from the fact that $g\in\mathcal{G}^H_2$, and thus $\E_{x'\in H}[ P_g'(S|x')]\leq 8$. Equation \ref{eq:PKb03} follows from the definition of $P'_g$. Equation \ref{eq:PKb04} follows from the definition of $\delta_g$. Equation \ref{eq:PKb1} follows from $g\in\mathcal{G}_1^H$ and thus $g(x_g)=\#H-2$.\\

\noindent\textbf{(7).}  \textit{This section shows that there is a $g\in\mathcal{G}^X_1\cap \mathcal{G}^X_2$, where $X$ is from the statement of the theorem, such that $\K(g|b)\lel \I(X:\mathcal{H}|b)$. We use a finite set $D\subset\mathcal{G}$, and since it has large $\kappa$ measure, using Theorem \ref{thr:el}, the complexity of $\K(g|b)$ is bounded. The desired results are achieved using Lemmas \ref{lem:StochTotalF} and \ref{lem:StochH}.}\\

\noindent Let $\{G_i\}$ be a computable enumeration of all finite subsets of $\mathcal{G}$. Let $f$ be a function that when given a set $H\subset\FS$, $\# H>2$, outputs an encoding of the first finite subset $W\subset\mathcal{G}$ in the list $\{G_i\}$ such that $W\subset\mathcal{G}^H_1\cap\mathcal{G}^H_2$ and $\kappa(W)> 0.25$. On all other inputs which are not an encoding of a finite set $H\subset\FS$ with $\#H>2$,  $f$ outputs the empty string. The function $f$ is total computable relative to $b$, with $\K(f|b)=O(1)$, because given $H$ and $b$, it is computable to determine whether a given function $g\in\mathcal{G}$ is in $\mathcal{G}_1^H\cap\mathcal{G}_2^H$.

Let $D=f(X)$.  Invoking Theorem \ref{thr:el}, conditional to $b$, gives $g\in D$, where $\K(g|b)\lea \ceil{-\log\m(D|b)}+2\K(\ceil{-\log\m(D|b)})+\Ks(D|b)$. Since $\ceil{-\log\m(D|b)}\lea -\log\kappa(D)+\K(\kappa|b)< O(1)$, we have that $\K(g|b)\lea\Ks(D|b)$. Lemma \ref{lem:StochTotalF}, relativized to $b$, using total computable function $f$, gives $\K(g|b)\lea\Ks(X|b)$. Lemma \ref{lem:StochH}, gives
\begin{align}
\label{eq:gFuncH}
\K(g|b)\lel \I(X:\ch|b). 
\end{align}

\noindent\textbf{(8).}  \textit{Using the inequality of Section 6, when $H=X$, an inequality can be derived about $-\log P_g$ using the terms $l$ and $k$. This results in the (general) inequality $\K_g\lessapprox-\log P_g +\K(g)\lessapprox l-k+\K(g)$, which is used to bound the expectation of $\K(x_g|x')$.}\\

\noindent Since $g\in D\subset \mathcal{G}_1^X\cap \mathcal{G}_2^X$, there exists $x_g\in X$ where, due to Equation \ref{eq:PKb1},
\begin{align}
\label{eq:Pglk}
\E_{x'\in X}[-\log P_g(x_g|x')]&\lel\max\{\E_{x'\in X}[\K[b](x_g|x')]-\#X,O(1)\}\lel l-k.
\end{align}
So we have that
\begin{align}
\nonumber
\E_{x'\in X}[\K(x_g|b,x')] &\lea \E_{x'\in X}[\K(x_g|b,g,x')+\K(g|b)]\\
\nonumber
&\eqa  \E_{x'\in X}[\K(x_g|b,g,x')]+\K(g|b)\\
\label{eq:lastEq1}
&<   \E_{x'\in X}[-\log P_g(x_g|x')]+\I(X:\ch|b)+O(\log\I(X:\ch|b))\\
\label{eq:lastEq2}
&< l-k+\I(X:\ch|b)+O(\log \I(X:\ch|b)+\log (l-k))\\
\label{eq:lastEq15}
\E_{x'\in X}[\K(x_g|x')-\K(b)] &< l-k+\I(X:\ch|b)+O(\log \I(X:\ch|b)+\log (l-k))\\
\nonumber
\E_{x'\in X}[\K(x_g|x')] &< l-k+\K(b)+\I(X:\ch|b)+O(\log (\I(X:\ch|b)+\K(b))+\log (l-k))
\end{align}
Equation \ref{eq:lastEq1} is due to Equation \ref{eq:gFuncH}. 
Equation \ref{eq:lastEq2} is due to Equation \ref{eq:Pglk}. Equation \ref{eq:lastEq15} follows that for all $x'\in X$, $\K(x_g|x')\lea \K(x_g|b,x')+\K(b)$.\\

\noindent\textbf{(9).} \textit{Using Lemma \ref{lmm:totalString}, the total string $b$ is removed from the inequality.} 
\begin{align}
\label{eq:lastEq3}
\E_{x'\in X}[\K(x_g|x')] &\lel l-k+\I(X:\ch)+\K(b|X)\\
\label{eq:lastEq4}
\E_{x'\in X}[\K(x_g|x')] &\lel l-k+\I(X:\ch)+\K(\langle\|b\|,(l-k)\rangle).\\
\label{eq:lastEq5}
\E_{x'\in X}[\K(x_g|x')] &\lel l-k+\I(X:\ch).
\end{align} Equation \ref{eq:lastEq3} is due to the invocation of Lemma \ref{lmm:totalString}. Equation \ref{eq:lastEq4} is due to the fact that $\K(b|X)\lea\K(\langle\|b\|,(l-k)\rangle)$. Equation \ref{eq:lastEq5} is because $a\lel b+O(\log a)$ implies $a\lel b$, where $a=\|b\|\lea \K(b)$ and $b=\I(X:\ch)+O(\log \|b\|)$. \qed
	\end{prf}
\section{Bunches}
\label{sec:bunches}
Recall that a $(k,l)$ bunch $X$ is a finite set of strings, where $k=\# X$, $l>k$, and for all $x,x'\in X$, $\K(x|x')\leq l$. The following theorem states that for non-exotic bunches, there is an element of $X$ that is simple conditional to all other members of $X$.
\begin{thr}
	\label{thr:bunches}	For $(k,l)$ bunch $X$, $\min_{x\in X}\max_{x'\in X}\K(x|x')\lel 2(l-k)+\I(X:\ch)$.
\end{thr}

\noindent\textbf{Informal Proof.}\\\

\noindent \textit{This proof starts with the definition of elementary probability measure $Q$ that realizes the stochasticity of $X$. Using probabilistic arguments, we define a $Q$-test $t_g$ that gives a high score to a set $Y$ if there does not exist $a\in Y$ such that $g(a)\gtrapprox \# Y$. A measure is defined by $P_g(x|y)\approx [g(x)\geq\K[b](x|y)-z]2^{-z}$, where $z=l-k$. A second test $t'_g$ gives a set $Y$ a zero score if more than half of $x'\in Y$ makes $P_g(\cdot|x')$ a semi-measure.  By probabilistic arguments, there exists a function $g$ such that $t_g$ and $t'_g$ are $Q$-tests. Furthermore, since $X$ is typical of $Q$, $t_g(X)=t'_g(X)=0$. Thus there exist $x_g\in X$ where $g(x_g)\geq \# X\geq\K[b](x|y)-z$, for all $x\in X$..  This means that $P_g(x_g|y)\approx 2^{-z}$ for all $y\in X$. By the fact that $t'(X)=0$, for more than half $x'\in X'\subseteq X$, $P_g(\cdot|x')$ is a semimeasure, and thus $\K(x_g|x')\lessapprox -\log P_g(x_g|x')$. For $x'\in X'$, the bound of theorem is achieved. For $y'\in X\setminus X'$, there exists $\approx 2^{k}$ programs from $y'$ to $y\in X'$, and then there is a short program from $y$ to $x_g$ using $P_g$. Thus the algorithmic probability of $\m(x_g|y')$ is large and the bounds for $y'\in X\setminus X'$ is achieved. The remainder of the proof uses Lemma \ref{lem:StochH} to replace stochasticity with mutual information with the halting sequence and Lemma \ref{lmm:totalString} to remove the total string $b$.}
\begin{prf}$ $\\
	
\noindent\textbf{(1.)} \textit{The first step of the proof is to find a total string $b$ such that $X$ is a bunch with computable complexity $\K[b]$, with $\max_{x,x'\in X}\K[b](x|x')\lea l$. This enables the proof to move forward with computable complexity and probability. The total string $b$ is factored out at the end of the proof. In this section, the probability measure $Q$ that realizes the stochasticity of $X$ is defined.}\\ 

\noindent Let $z=l-k$ and let $b$ be the shortest total string where $\max_{x,x'\in X}\K[b](x|x')<l+2$, which we call satisfying property A. Thus $\K(b|X)\lea\K(\langle z,\|b\|\rangle)$ and $b^-$ is not total, using arguments in the first paragraph of the proof of Theorem \ref{thr:batch}. Let $s=\langle b,z\rangle$. Let $v\in \FS$ and elementary probability measure $Q$ minimize $\Ks(X|s)$, where $U_s(v)=Q$. Recall that elementary measures are introduced in Definition \ref{def:Elem}. Let  $d=\max\{\d(X|Q,\langle v,s\rangle),1\}$. Let $S=\bigcup\{Y:\langle Y\rangle \in\supp(Q)\}$ be the union of all sets encoded in the support of $Q$.  Since $Q$ is elementary, $|S|<\infty$.  Let $\mathcal{G}$ be the set of all functions $g:S\rightarrow\N$. Since $S$ is finite, each $g\in\mathcal{G}$ can be encoded with an explicit finite string.\\

\noindent\textbf{(2).} \textit{We define a probability measure $\kappa$ over functions $g\in\mathcal{G}$ from the union of the support of $Q$ to natural numbers, where functions with low values will have a higher probability.}\\

\noindent Let $\kappa : \mathcal{G}\rightarrow\R_{\geq 0}$ be a probability measure over $\mathcal{G}$, where $\kappa(g)=\prod_{a\in S}2^{-g(a)}$. So for all $a\in S$, $\kappa(\{g:g(a)\geq n\})=2^{-n+1}$. Let $c\in\N$ be a constant solely dependent on $U$ to be determined later. \\

\noindent\textbf{(3.)} \textit{The proof only works with $X$ having a minimum number of elements. Otherwise the theorem is trivially solved. This is a boundary case that can be skipped on first reading.}\\

\noindent We assume that $|X| > 16(c+d)$. Otherwise, $k \lea \log d$, and then $\min_{x\in X}\max_{x'\in X}\K(x|x',s)\leq l\lea z+\log d \lea 2z + \Ks(X|s)$. From this point, the reasoning starting at Equation \ref{eq:bunchlast1} can be used to prove the theorem. \\

\noindent\textbf{(4.)} \textit{We define the first of two tests, $t_g$, parameterized by a function $g\in\mathcal{G}$. We will show later in the proof there is a $g$ such that $t_g$ is a $Q$ test. $t_g$ gives a high score to sets $Y$ such that all their elements $a\in Y$ have low $g$ score.}\\

\noindent We define the following function $t$ over $\supp(Q)$, parameterized by $g\in\mathcal{G}$. Let $\mathcal{B}$ be the set of sets $G$ such that for all $x,x'\in G$, $\K[b](x|x')< \#G+z+2$.  Let $t_g(Y)=e^{2(d+c)-1}$ if $Y\cap\{a:g(a)\geq\floor{\log (|Y|/(c+d))}\}=\emptyset$ and $Y\in \mathcal{B}$, otherwise $t_g(Y)=0$. \\

\noindent\textbf{(5.)} \textit{Using probabilistic arguments, it is shown that the expectation of $t_g$ over $Q$ and $\kappa$ is small. This is required for probabilistic arguments to show the existence of a $g\in\mathcal{G}$ with $t_g$ being a $Q$-test.}\\

\noindent So, using the fact that $(1-m)e^m\leq 1$,
\begin{align*}
&\;\;\;\;\;\E_{g\sim\kappa}\E_{Y\sim Q}[t_g(Y)]\\
 &= \sum_{Y\in\mathcal{B}}Q(Y)\kappa(\{g:\forall_{a\in Y}, g(a)<\floor{\log (|Y|/(c+d))}\})e^{2(d+c)-1}\\
&= \sum_{Y\in\mathcal{B}}Q(Y)\prod_{a\in Y}\kappa(\{g: g(a)<\floor{\log (|Y|/(c+d))}\})e^{2(d+c)-1}\\
&= \sum_{Y\in\mathcal{B}}Q(Y)\prod_{a\in Y}\left(1-2^{-\floor{\log (|Y|/(c+d))}+1}\right)e^{2(d+c)-1}\\
&\leq \sum_{Y\in\mathcal{B}}Q(Y)(1-2(c+d)/|Y|)^{|Y|}e^{2(d+c)-1}\\
&\leq \sum_{Y\in\mathcal{B}}Q(Y)e^{-2(c+d)}e^{2(c+d)-1}< 0.5.\\
\end{align*}

\noindent\textbf{(6.)} \textit{The measure $P_g$ is defined, parameterized by $g\in\mathcal{G}$ gives $P_g(x|y)$ a score of $\approx 2^{-z}$ if $g(x)\gtrapprox \K[b](x|y)-z$ and 0 otherwise. The constants and max function ensure proper boundary conditions and can be discounted on a first reading. By definition, the expection of $P_g(S|y)$, over $g$ distributed by $\kappa$ is small.}\\

\noindent For each $x,y\in S$, $g\in\mathcal{G}$, we define the following function $$P_g(x|y)=[g(x)\geq\max\{ \K[b](x|y)-z-\ceil{\log (c+d)}-3,1\}]2^{-z-2(d+c)}.$$
Thus $P_g(x|y)$ is only one of two values, either 0 or $2^{-z-2(d+c)}$. $P_g(S|y)=\sum_{x\in S}P_g(x|y)$. So for all $y\in S$, we have 
\begin{align*}
&\;\;\;\;\E_{g\sim \kappa}[P_g(S|y)] \\
&= 2^{-z-2(d+c)}\sum_{x\in S}\kappa(\{g:g(x)\geq \max\{\K[b](x|y)-z-\ceil{\log (d+c)}-3,1\})\\
&= 2^{-z-2(d+c)}\sum_{x\in S}2^{-\max\{\K[b](x|y)-z-\ceil{\log(d+c)}-3,1\}+1}\\
&\leq 2^{-z-2(d+c)}\sum_{x\in S}\m[b](x|y)2^{z+\ceil{\log(d+c)}+4}\\
&\leq 2^{-(d+c)+4}.
\end{align*}

\noindent\textbf{(7.)} \textit{We define an indicator function $\I_g(y)$ which is 0 iff $P_g(\cdot|y)$ is a semi-measure, and $\I_g(Y)$ counts the number of non semi-measures using $y\in Y$. Using bounds of the previous section, an upper bound on the expectaction of $\I_g$ is given.}\\

\noindent For all functions $g\in \mathcal{G}$, we define the following indicator function, with $\I_g(y)=[P_g(S|y)>1]$. Furthermore, we extend the domain $\I$ to be over sets $Y\in\supp(Q)$, with $\I_g(Y)=\sum_{y\in Y}\I_g(y)$. Thus $\I_g(y)=0$ iff $P_g(\cdot|y)$ is a semimeasure where each $x\in \supp(P_g(\cdot|y))$ can be identified by code of size $\eqa-\log P_g(x|y)$. For each such $y\in S$, the expectation of $\I$ with respect to $\kappa$ is small, and for $Y\in\supp(Q)$, we have
\begin{align*} 
\E_{g\sim\kappa}[\I_g(y)]&\leq \E_{g\sim \kappa}[P_g(S|y)]  \leq 2^{-(c+d)+4}\\
\E_{g\sim\kappa}[\I_g(Y)]&\leq |Y|2^{-(c+d)+4}.
\end{align*}

\noindent\textbf{(8.)} \textit{We define the second test function $t'$, parameterized by $g\in \mathcal{G}$. It gives a set a zero score if $P_g$ is a semi-measure for at least half its elements. Otherwise it gives the set a high score. Through probabilistic arguments $t'_g$ has low $Q$ expectation when $g$ is distributed by $\kappa$. Note that since $\E_{g\sim\kappa}[\I_g(Y)]\leq |Y|2^{-(c+d)+4}$, by the Markov inequality $\kappa(g:\I_g(Y)\geq 0.5|Y|)\leq2^{-(c+d)+5}$.}\\

\noindent We define the function $t':\supp(Q)\rightarrow \R_{\geq 0}$, parameterized by $g\in\mathcal{G}$, which will give a set $Y$ a zero score iff $P_g(\cdot|y)$ is a semi-measure for at least half of the elements $y\in Y$. Otherwise $t'_g$ gives $Y$ a high score. More formally, let $t'_g(Y)=0$ if $\I_g(Y)<.5|Y|$ and $t'_g(Y)=2^{(d+c)-7}$, otherwise. Thus we have that, due to the Markov inequality,
\begin{align*}
\E_{g\sim\kappa}\E_{Y\sim Q}[t'_g(Y)] &=\sum_YQ(Y)\E_{g\sim\kappa}[[g:\I_g(Y)\geq 0.5|Y|]]2^{c+d-7}\\
 &=\sum_YQ(Y)\kappa(\{g:\I_g(Y)\geq 0.5|Y|\})2^{c+d-7}\\
 &\leq\sum_YQ(Y)2^{-(c+d)+5}2^{c+d-7}\\
 &=0.25.
\end{align*}

\noindent\textbf{(9.)} \textit{Since the $\kappa$-expectation of $t_g$ and $t_g'$ are small, by probabilistic arguments, there is a $g\in\mathcal{G}$ where $t_g$ and $t_g'$ are both $Q$-tests. Using similar arguments to that in the proof of Theorem \ref{thr:batch}, it is proven that $t_g(X)=t'_g(X)=0$.}\\

\noindent By probabilistic arguments, there exists $g\in \mathcal{G}$, such that $\E_{Y\sim Q}[t_g(Y)]\leq 1$ and  $\E_{Y\sim Q}[t'_g(Y)]\leq 1$. So both $t_g(\cdot)Q(\cdot)$ and $t'_g(\cdot)Q(\cdot)$ are semi-measures. Furthermore, $\K(g|c,d,v,s) = O(1)$. It must be that $t_g(X)=0$. Otherwise, for proper choice of $c$ solely dependent on $U$,
\begin{align*}
d &=\d(X|Q,v,s)\\
&= \ceil{-\log Q(X)}-\K(X|v,s)\\
& > -\log Q(X)-(-\log t_g(X)Q(X)+\K(t_g(\cdot)Q(\cdot)|v,s))-O(1)\\
& > -\log Q(X)-(-\log t_g(X)Q(X)+\K(g,Q|v,s))-O(1)\\
& > 2(c+d)(\log e)-\K(c,d)-O(1)\\
&> d,
\end{align*}
causing a contradiction. Thus $c$ is chosen to be large enough to have the property $c>\K(c)+O(1)$, where the additive constant is depedent solely on the universal Turing machine. The same reasoning can be used to show that $t'_g(X)=0$. We roll $c$ into the additive constants of the theorem and remove it from consideration for the rest of the proof.\\

\noindent\textbf{(10.)} \textit{Since $t_g(X)=0$, there exists $a\in X$ where $g(a)$ has a high score, with $g(a) \gtrapprox \K[b](a|y)-z$, for all $y\in X$. The inequality follows from $k\geq k+\K[b](a|y)-l= \K[b](a|y)-z$. This ensures that $P_g(a|y)\approx 2^{-z}$ for all $y\in X$.}\\

\noindent Therefore, since $t_g(X)= 0$, there exists $a\in X$ where for all $y\in X$, using the fact that $|Y| > 16(c+d)$,
\begin{align*}
g(a)&\geq \floor{\log (|Y|/(d+c)}\\
&\geq \floor{\log |Y|}-\ceil{\log(d+c)}\\
&\geq k-1-\ceil{\log(d+c)}\\
&\geq \max\{\K[b](a|y)-z-\ceil{\log(d+c)}-3,1\}.
\end{align*}
This ensures that $P_g(a|y)>0$ for all $y\in X$, due to the definition of $P_g$.\\

\noindent\textbf{(11.)} \textit{Since $t'_g(X)=0$, $P_g(\cdot|y)$ is a semimeasure  for more than half $X'$ of $y\in X$. Thus $P_g(a|y)$ can be used to identify $a$ given $y$ in this subset $X'$ and the desired bound on $\K(a|y)$ is achieved. Otherwise for $y'\not\in X'$, a program can be created that computes some $y\in X'$ from $y'$ (bounded by $l$) and then use the bound proved of $\K(a|y)$. Since there is a lot of $y\in X'$, there is a lot of such programs, meaning the algorithmic probability of $\m(a|y')$ is large, and thus the bound is achieved.}\\

\noindent Furthermore, since $t'_g(X)= 0$, there is a subset $X'\subseteq X$, $|X'|> 2^{k-2}$, where for all $y\in X'$, $P_g(\cdot|y)$ is a semimeasure. For such $y$, $\K(a|y,s)\lea -\log P_g(a|y)+\K(g|d,v,s)+\K(d,v|s)\lea z+3d + \|v\|\lea z+\Ks(X|s)$. Therefore for all $y'\in X\setminus X'$, 
\begin{align*}
\K(a|y',s) &\lea -\log \sum_{y\in X'}2^{-\K(a|y,s)-\K(y|y',s)}\\
	& \lea -\log \sum_{y\in X'}2^{-l-z-\Ks(X|s)}\\
	&\lea 2z+\Ks(X|s).
\end{align*}
\noindent\textbf{(12.)} \textit{The following theorem removes the stochasticity term and the total string $b$, similarly to the proof of Theorem \ref{thr:batch}.}\\

\noindent So for all $x\in X$, 
\begin{align}
     \label{eq:bunchlast1}
\K(a|x,s) &\lea 2z+\Ks(X|s)\\
\nonumber
\K(a|x) & \lea 2z +\K(s)+\Ks(X|s)\\
\nonumber
     & < 2z +\K(b)+\Ks(X|s)+O(\log z)\\
     \label{eq:bunchlast2}
     & < 2z +\K(b)+\I(X:\mathcal{H}|s)+O(\log z+\log \I(X:\mathcal{H}|s))\\
     \nonumber
     & < 2z +\K(b)+\I(X:\mathcal{H}|b)+O(\log z+\log (\I(X:\mathcal{H}|b)+\K(b)))\\
     \label{eq:bunchlast3}
     & \lel 2z+\I(X:\mathcal{H})+\K(b|X)\\
      \nonumber
     & \lel 2z+\I(X:\mathcal{H})+\K(\langle \|b\|,z\rangle)\\
     \label{eq:bunchlast4}
     & \lel 2z+\I(X:\mathcal{H}).
\end{align}
Equation \ref{eq:bunchlast2} is due to the application of Lemma \ref{lem:StochH}. Equation \ref{eq:bunchlast3} is due to the application of Lemma \ref{lmm:totalString}. Equation \ref{eq:bunchlast4} uses the same logic as Equation \ref{eq:lastEq5} in the proof of Theorem \ref{thr:batch}.\qed
\end{prf}

\section{Discussion}
There exists a generalization for batches to elementary probability measures, where for batch $X$, $l \geq \max_{x\in X}\E_{x'\sim X}[\K(x|x')]$ and $k=\#\supp(X)$. For bunches, there is a way to achieve a comparable result to Theorem \ref{thr:bunches}, for enumerative sets $X$, where instead of the bounds being in terms of $\I(X:\mathcal{H})$, the bounds are in terms of $\I(p:\ch)$, where $p$ is a program that enumerates $X$. In both cases, we leave the details of the proofs to the reader. 

The stochasticity method has been proven fruitful in characterizing elementary objects that have low mutual information with the halting sequence. Further work involves publishing results regarding stochasticity and the $\M$ measure of prefix free sets, where $\M$ is the universal lower computable continuous semi-measure. This work has application in the minimal complexity of completions of partial binary predicates. Other work involves looking at stochasticity and combinatorial objects, such as graphs or matroids.


\end{document}